\documentclass[sigconf]{acmart} 

\usepackage{tikz}
\usetikzlibrary{positioning,arrows.meta,calc}
\usepackage{pifont}  
\usepackage{multirow}
\usepackage{amsmath}
\usepackage{algpseudocode}
\usepackage{graphicx}
\usepackage{textcomp}
\usepackage{color}
\usepackage{xcolor}
\usepackage{booktabs}
\usepackage{xspace}
\usepackage{tcolorbox}
\usepackage{algorithm}
\usepackage{mathtools}
\usepackage{array}
\usepackage{soul}
\usepackage{float}
\usepackage{caption}
\usepackage{subcaption}
\usepackage{enumitem}
\usepackage{balance}
\usepackage[table,dvipsnames]{xcolor}
\usepackage{booktabs}
\usepackage{tabularx}
\usepackage{siunitx}
\newcommand{\TheName}{\textsc{Agent4Decompile}\xspace}
\newcommand{\projname}{\textsc{Agent4Decompile}\xspace}

\newcommand{\finding}[2]{%
\vspace{0.5em}
\noindent\fbox{\parbox{0.95\columnwidth}{\textbf{Finding #1:} #2}}
\vspace{0.5em}
}

\newcommand{\squishlist}{
 \begin{list}{$\bullet$}
   { \setlength{\itemsep}{0pt}
     \setlength{\parsep}{0pt}
     \setlength{\topsep}{2pt}
     \setlength{\partopsep}{0pt}
     \setlength{\leftmargin}{1.5em}
     \setlength{\labelwidth}{1em}
     \setlength{\labelsep}{0.5em} } }
\newcommand{\squishend}{\end{list}}

\begin{document}

\date{}

\title{Constraint-Guided Multi-Agent Decompilation for Executable Binary Recovery}


\author{
{\rm Yifan Zhang$^{*}$, \quad Xiaohan Wang$^{*}$, \quad Yueke Zhang, 
\quad Yu Huang,
\quad Kevin Leach$^{\dagger}$} \\
Vanderbilt University \\
\normalsize $^{*}$Equal contribution.\quad $^{\dagger}$Corresponding author: \texttt{kevin.leach@vanderbilt.edu}
}

\pagestyle{plain}


\begin{abstract}
Decompilation---recovering source code from compiled binaries---is essential for many software engineering tasks, like program analysis, reverse engineering, and legacy software maintenance.
However, existing decompilers produce code that often fails to \emph{re}compile or execute correctly, limiting their practical utility in settings where changes to decompiled source must be lowered again to binary form.
Recent advances in agentic AI offer a promising direction: by defining structured workflows with clear validation signals, LLM agents can iteratively refine code toward correctness without task-specific training.
In this paper, we present \TheName{}, a multi-agent framework that transforms decompiled code into re-executable source through multi-level constraint-guided refinement.
Our approach operates in three stages:
(1)~\textit{syntax validation} ensures parseable C code;
(2)~\textit{compilation validation} catches type errors and missing declarations;
(3)~\textit{execution validation} uses test-driven behavioral equivalence against the original binary.
When validation fails at any level, specialized LLM agents refine the code using structured error feedback.
We conduct the largest decompilation experiment to date, evaluating \TheName{} on 1,641 binaries from ExeBench across three decompilers (Ghidra, Angr, RetDec)---a benchmark that enables fine-grained analysis across optimization levels, function categories, and decompiler architectures.
\TheName{} achieves 40--46\% re-executability on this large-scale benchmark, improving baseline output by 18--28 percentage points.
Compared to state-of-the-art methods, \TheName{} (50.3\%) outperforms single-pass LLM refinement (35.2\%) and fine-tuned LLM4Decompile (12.1\%) by 15--38 percentage points.
Our ablation study reveals that execution-based validation is critical: compile-only approaches achieve only 32--42\% re-executability despite 99--100\% compilation rates.
Moreover, \TheName{} converges efficiently, with most binaries reaching correctness within 2 iterations.
\end{abstract}

\maketitle


\section{Introduction}
\label{sec:intro}

Decompilation is the process of recovering high-level source code from compiled binary executables.
This challenging task is a linchpin for many software security and reverse engineering tasks. 
Security analysts rely on decompilation for malware analysis~\cite{yakdan2015no}, vulnerability discovery~\cite{pewny2015cross}, and understanding proprietary or legacy software~\cite{cifuentes1994reverse}.
An ideal decompiler would produce source code that is not only readable but also \textit{re-compilable} and \textit{re-executable} --- that is, capable of being compiled and executed to produce the same behavior as the original binary.
Such re-executable decompilation would enable powerful downstream applications including binary rewriting, automated patching, and fuzz testing of source code recovered via decompilation.

Unfortunately, existing decompilers fall far short of this ideal.
State-of-the-art tools like Ghidra~\cite{ghidra}, IDA Pro~\cite{hex2018ida}, and RetDec~\cite{retdec} employ sophisticated program analysis techniques, but produce output that frequently contains syntactic errors, type mismatches, and semantic inconsistencies.
Our empirical analysis reveals that raw decompiler output achieves only 22--26\% re-executability across different tools (Table~\ref{tab:main-results}).
This gap between decompiler output and usable source code represents a substantial barrier to practical binary analysis workflows.

Recent advances in Large Language Models (LLMs) have sparked interest in leveraging neural approaches for decompilation~\cite{llm4decompile,sk2decompile}.
These methods treat decompilation as a translation task, training or prompting LLMs to convert assembly or decompiled pseudocode into compilable C code.
While promising, existing LLM-based approaches face fundamental limitations: they operate in a single-shot manner without feedback, lack mechanisms to verify behavioral correctness.
More fundamentally, these limitations suggest that decompilation cannot be solved as a single-pass translation problem.

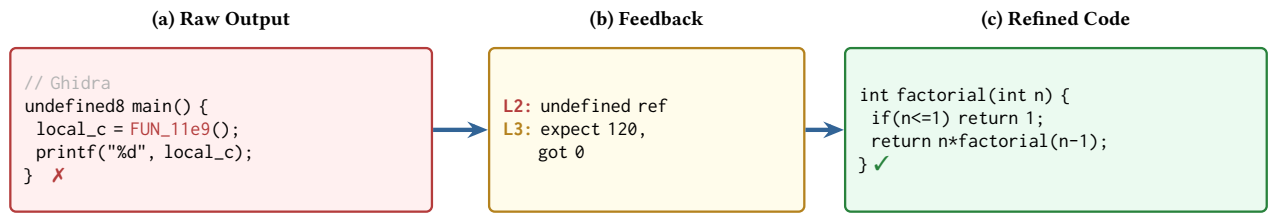
\begin{figure*}[t]
    \centering
    \resizebox{0.95\textwidth}{!}{%
\begin{tikzpicture}[
    every node/.style={font=\footnotesize},
    codebox/.style={draw, thick, rounded corners=2pt, text width=4.8cm, align=left, inner sep=5pt, font=\ttfamily\footnotesize, minimum height=2.0cm},
    labelbox/.style={font=\footnotesize\bfseries},
    bigarrow/.style={-{Stealth[length=2.5mm, width=2mm]}, line width=1.5pt}
]

\definecolor{rawbg}{RGB}{255,240,240}
\definecolor{rawborder}{RGB}{180,60,60}
\definecolor{errorbg}{RGB}{255,252,235}
\definecolor{errorborder}{RGB}{180,130,30}
\definecolor{goodbg}{RGB}{235,250,240}
\definecolor{goodborder}{RGB}{40,130,60}
\definecolor{arrowcolor}{RGB}{50,100,150}

\node[codebox, fill=rawbg, draw=rawborder] (raw) at (0, 0) {%
\textcolor{gray!60}{// Ghidra}\par
undefined8 main() \{\par
\hspace{4pt}local\_c = \textcolor{rawborder}{FUN\_11e9}();\par
\hspace{4pt}printf("\%d", local\_c);\par
\}\quad\textcolor{rawborder}{\ding{55}}%
};
\node[labelbox, above=0.1cm of raw] {(a) Raw Output};

\node[codebox, fill=errorbg, draw=errorborder, text width=3.5cm] (error) at (5.2, 0) {%
\textcolor{rawborder}{\textbf{L2:}} undefined ref\par
\textcolor{errorborder}{\textbf{L3:}} expect 120,\par
\hspace{12pt}got 0%
};
\node[labelbox, above=0.1cm of error] {(b) Feedback};

\node[codebox, fill=goodbg, draw=goodborder] (refined) at (10.2, 0) {%
int factorial(int n) \{\par
\hspace{4pt}if(n<=1) return 1;\par
\hspace{4pt}return n*factorial(n-1);\par
\} \textcolor{goodborder}{\ding{51}}%
};
\node[labelbox, above=0.1cm of refined] {(c) Refined Code};

\draw[bigarrow, arrowcolor] (raw.east) -- (error.west);
\draw[bigarrow, arrowcolor] (error.east) -- (refined.west);

\end{tikzpicture}
    }
    \caption{Motivating example: (a) Raw Ghidra output contains undefined functions and type errors. (b) Multi-level constraints provide structured feedback: L2 catches missing function references, L3 catches behavioral errors. (c) \TheName{} produces correct, re-executable code through iterative refinement.}
    \label{fig:example}
\end{figure*}

In this paper, we present \TheName{}, a multi-agent framework that transforms decompiled code into re-executable source through multi-level constraint-guided refinement.
Our key insight is that decompilation is inherently a multi-level problem arising from the lossy nature of compilation.
During compilation, code is lowered to assembly and machine code, leading to a loss different types of information at different stages.
Consequently, no single validation signal can recover all lost information.
Variable and struct names are discarded early, causing syntax-level inconsistencies in decompiled output; type information is erased, leading to compilation failures; and control flow semantics are obscured by optimization, resulting in behavioral mismatches.
Crucially, these error classes are \emph{not jointly optimizable}: execution feedback cannot fix syntax errors, and compilation diagnostics cannot detect semantic mismatches.
This necessitates a hierarchical validation strategy that integrates complementary signals across levels.

Therefore, we formulate decompilation refinement as a \emph{progressive search space reduction}: given the vast space of possible C programs, each constraint level defines a successively smaller feasible subspace---L1 (syntax) restricts to parseable programs, L2 (compilation) further narrows to type-correct programs, and L3 (execution) identifies behaviorally equivalent programs.
Figure~\ref{fig:example} illustrates this progression.
By decomposing the search into hierarchical stages, \TheName{} makes the refinement tractable: rather than searching the entire program space with a single weak signal, we progressively filter through increasingly precise constraints, each providing targeted feedback for its corresponding error class.

Our ablation study reveals a critical finding: \textit{compilation success does not imply behavioral correctness}.
A compile-only validation approach (Levels 1+2) achieves 99--100\% compilation rates but only 32--42\% re-executability; the code compiles but produces incorrect outputs.
Incorporating execution-based validation (Level 3) improves re-executability to 43--50\%, closing part of this gap.
This result underscores the necessity of end-to-end validation for re-executable decompilation.

On a 157-binary benchmark, \TheName{} improves re-executability from 22--26\% (baseline) to 43--50\% across three decompilers (Ghidra, Angr, and RetDec), representing improvements of 18--28 percentage points.
Compared to state-of-the-art refinement methods, \TheName{} (50.3\%) outperforms single-pass LLM refinement (35.2\%) and fine-tuned LLM4Decompile-Ref (12.1\%) by 15--38 percentage points.
The system scales effectively: on 1,641 binaries from ExeBench, \TheName{} maintains 40--46\% re-executability with an average of 1.5--2.3 iterations per binary.

This paper makes the following contributions:

\begin{enumerate}[leftmargin=*,topsep=2pt,itemsep=1pt]
\item \textbf{Insight:} We identify that compilation success does not imply behavioral correctness: a 57--68 pp gap exists between compile rate (99\%) and re-executability (32--42\%).  This motivates multi-level validation with execution feedback (\S\ref{sec:design}).

\item \textbf{System:} We present \TheName{}, a multi-agent framework that iteratively refines decompiled code through syntax, compilation, and execution constraints, improving re-executability from 22--26\% to 43--50\% across three decompilers (\S\ref{sec:implementation}).

\item \textbf{Evaluation:} We demonstrate 15--38 pp improvement over prior refinement methods on 1,641 binaries, with ablations showing L3 execution feedback is essential for behavioral correctness (\S\ref{sec:eval}). We release all benchmarks and artifacts at: \url{https://anonymous.4open.science/r/agent4decompile-artifacts-0F69}.
\end{enumerate}


\section{Background}
\label{sec:background}

We provide background on binary decompilation and LLMs for code understanding, which form the foundation of our approach.

\subsection{Binary Decompilation}
\label{sec:background:decompilation}

Decompilation converts compiled binary executables back into human-readable source code.
Unlike disassembly (which produces assembly from binaries), decompilation aims to recover high-level constructs such as control flow statements, data types, and variable names in a higher level language (typically C).
Modern decompilers follow a multi-stage pipeline~\cite{yakdan2015no}: binary parsing loads the executable and identifies code/data sections; disassembly converts machine code to assembly instructions; control flow recovery reconstructs the control flow graph (CFG); type recovery infers variable types from usage patterns; and code generation produces source code from the intermediate representation.

Three factors make decompilation fundamentally challenging~\cite{yakdan2015no}.
First, compilation is an inherently lossy transformation: information such as variable names, comments, and high-level abstractions are irreversibly lost as high level source code is lowered to assembly and then to machine code.
Second, compiler optimizations (inlining, loop unrolling, dead code elimination) obscure the program's original structure or the intent of the original developer. 
Third, the diversity of calling conventions, instruction set architectures, and compiler behaviors creates a vast space of possible binary patterns.

Modern decompilers fall into three architectural categories:
\begin{itemize}[leftmargin=*,nosep]
    \item \textbf{Rule-based decompilers} (e.g., Ghidra~\cite{ghidra}, IDA Pro~\cite{hex2018ida}) use hand-crafted pattern matching rules developed over decades to recognize code patterns and produce output.
    
    \item \textbf{Lifting-based decompilers} (e.g., Angr~\cite{shoshitaishvili2016sok}, RetDec~\cite{retdec}) translate binary code to an intermediate representation (IR) such as VEX or LLVM IR, then lower to source code.
    
    \item \textbf{ML-based decompilers} leverage machine learning to learn the mapping from binary to source code, either through neural machine translation~\cite{fu2019coda} or large language models~\cite{llm4decompile}.
\end{itemize}
Each architecture exhibits distinct error patterns: rule-based decompilers may miss modern compiler idioms; lifting-based decompilers can introduce IR artifacts; ML-based approaches may hallucinate or generate syntactically invalid code.

A fundamental question in decompilation is: when is the decompiled code correct?
Syntactic equivalence, where decompiled code is textually identical to the original source, is rarely achievable due to information loss.
Semantic equivalence, where decompiled code produces the same outputs as the original binary for all inputs, represents the ideal goal but is generally undecidable.
We therefore adopt \emph{behavioral equivalence} as our correctness criterion: we say the decompiled code is \emph{correct} if it compiles, executes, and produces outputs matching the original binary on a set of test inputs.
This \emph{re-executability} metric is stricter than mere compilation success and provides a practical approximation of semantic equivalence.

\subsection{LLMs for Code Understanding}
\label{sec:background:llm}

Given the challenges of traditional decompilation, recent work has turned to large language models as a complementary approach.
Models such as Codex~\cite{chen2021evaluating}, CodeLlama~\cite{roziere2023code}, and StarCoder~\cite{li2023starcoder} are trained on large corpora of source code and demonstrate remarkable capabilities in code generation~\cite{chen2021evaluating}, code summarization~\cite{ahmed2022few}, and bug detection~\cite{pearce2022asleep}.
These models capture both syntactic patterns and semantic relationships in code.

Recent work applies LLMs directly to decompilation. 
LLM4Decompile~\cite{llm4decompile} fine-tunes language models on assembly-to-C pairs, achieving state-of-the-art results on decompilation benchmarks. However, these approaches operate in a single-pass manner: given binary input, they produce source code without iterative refinement, leaving many errors uncorrected.

LLM-based agents extend language model capabilities by enabling interaction with external tools and environments~\cite{yao2022react,shinn2023reflexion}. Agents can execute code, observe outputs, and iteratively refine their solutions based on feedback. This paradigm has shown success in code generation~\cite{hong2023metagpt}, debugging~\cite{chen2023teaching}, and program repair~\cite{xia2023automated}. Our work builds on this agent paradigm, using execution feedback to guide iterative decompilation refinement.

\section{System Design}
\label{sec:design}

\begin{figure}[t]
    \centering
    \includegraphics[width=\columnwidth]{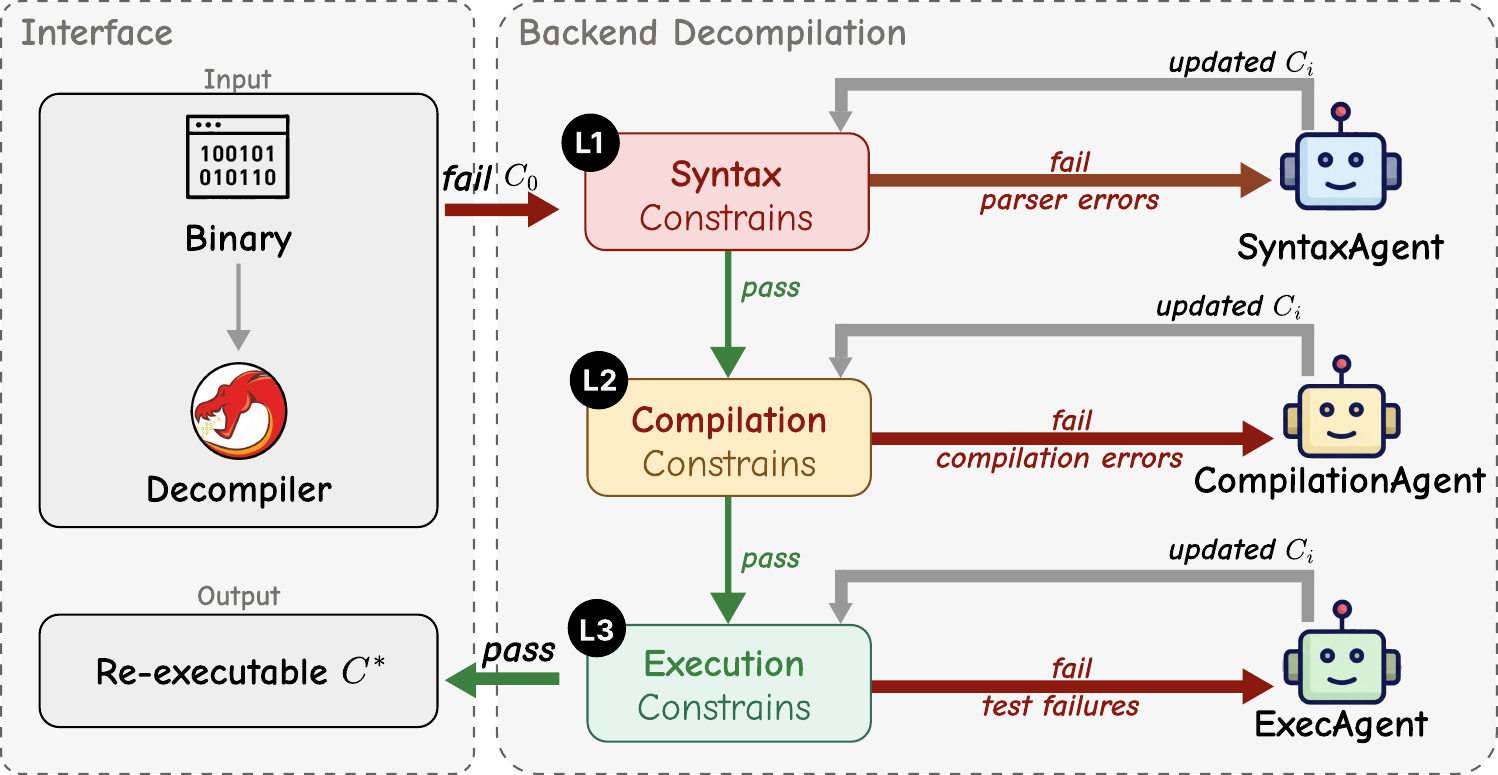}
    \caption{Overview of \projname{}. A binary is first processed by a traditional decompiler (e.g., Ghidra) to produce initial code $C_0$. The code then passes through a three-level constraint hierarchy: L1 (syntax), L2 (compilation), and L3 (execution). At each level, failures trigger a specialized LLM agent that repairs the code using error feedback. Code that passes all three levels yields re-executable output $C^*$.}
    \label{fig:overview}
\end{figure}

We introduce \projname{}, a system that transforms decompiled C code into re-executable source through constraint-guided LLM refinement.
Figure~\ref{fig:overview} illustrates our approach: given a binary, a traditional decompiler produces initial code $C_0$, which then passes through three levels of validation (syntax, compilation, execution), with LLM-based repair at each level until the output compiles and behaves equivalently to the original binary.

\subsection{Overview}
\label{sec:design:overview}

\paragraph{Scope and Assumptions.}
We target stripped ELF binaries compiled from C with standard optimization levels (O0--O3). We assume test inputs are available for behavioral validation, either from domain knowledge, fuzzing, or LLM-generated cases. We do not address obfuscated binaries or architectures beyond x86-64; these extensions are left to future work.

\paragraph{Pipeline.}
\projname{} operates as a four-stage pipeline: (1) decompilation using an off-the-shelf decompiler, (2) post-processing to normalize output, (3) optional multi-decompiler consensus, and (4) constraint-guided LLM refinement. This section focuses on the refinement stage, which is our main contribution.

Existing decompilers operate in a single pass: they produce output without any mechanism to detect or correct errors. \projname{} addresses this limitation by treating decompilation as an iterative process guided by executable constraints.

Our key insight is that \emph{decompilation errors manifest at different levels}, and each level provides distinct diagnostic signals:

\begin{itemize}[leftmargin=*,nosep]
    \item \textbf{L1: Syntax errors} (e.g., missing semicolons) indicate malformed code that cannot be parsed by the compiler front-end.
    \item \textbf{L2: Compilation errors} (e.g., type mismatches between declarations and uses of variables) indicate semantic inconsistencies.
    \item \textbf{L3: Execution errors} (e.g., wrong outputs) indicate behavioral divergence from the original binary.
\end{itemize}

The three levels form a validation pipeline: L1 checks syntax, L2 checks compilation, and L3 checks behavioral equivalence. When validation fails at any level, a specialized LLM agent receives the error feedback and proposes targeted repairs. Code passing all three levels yields re-executable output.

\subsection{Constraint Hierarchy}
\label{sec:design:constraints}

We first define the domains used in our formalization. Let $\mathcal{S}$ denote the set of all strings, $\mathcal{C} \subset \mathcal{S}$ the set of syntactically valid C programs, and $\mathcal{B}$ the set of executable binaries. We define the following operators:
\begin{itemize}[leftmargin=*,nosep]
    \item $\text{parse}: \mathcal{S} \to \mathcal{C} \cup \{\bot\}$ attempts to parse a string as C code.
    \item $\text{compile}: \mathcal{C} \to \mathcal{B} \cup \{\bot\}$ compiles valid C code to a binary.
    \item $\text{exec}: \mathcal{B} \times \mathcal{I} \to \mathcal{O}$ executes a binary on input $i \in \mathcal{I}$ and returns output $o \in \mathcal{O}$.
\end{itemize}

\paragraph{L1: Syntax Constraints.}
Given decompiled code $C \in \mathcal{S}$, the syntax constraint holds if $C$ is parseable:
\[
\phi_1(C) \triangleq \text{parse}(C) \neq \bot
\]
When $\phi_1(C) = \textsc{false}$, the parser returns error messages $E_1$ indicating the location and nature of syntax errors (e.g., \texttt{line 15: expected `;' before `\}'}). Separating syntax validation from compilation provides more localized feedback, enabling targeted repairs before attempting the more complex compilation step.

\paragraph{L2: Compilation Constraints.}
The compilation constraint holds if parsed code compiles successfully:
\[
\phi_2(C) \triangleq \phi_1(C) \land \text{compile}(\text{parse}(C)) \neq \bot
\]
When $\phi_2(C) = \textsc{false}$, the compiler returns error messages $E_2$ with type information and symbol resolution failures. Note that $\phi_2(C) \Rightarrow \phi_1(C)$: compilation subsumes parsing.

\paragraph{L3: Execution Constraints.}
Given a test suite $T = \{(i_1, o_1), \ldots, (i_n, o_n)\}$ derived from the original binary $B$, the execution constraint holds if the recompiled code produces matching outputs:
\[
\phi_3(C, T) \triangleq \phi_2(C) \land \forall (i, o) \in T: \text{exec}(\text{compile}(C), i) = o
\]
When $\phi_3(C, T) = \textsc{false}$, we obtain the failing test case $(i, o_{\text{exp}}, o_{\text{act}})$ as diagnostic feedback. This level validates behavioral equivalence---the ultimate goal of decompilation.

\subsection{LLM-Based Repair Agents}
\label{sec:design:agents}

\projname{} employs three specialized LLM agents, one for each constraint level. While the agents share a common prompt structure, each receives level-specific error feedback tailored to its repair task.

\paragraph{Prompt Structure.}
All agents receive prompts with four components:
\begin{enumerate}[leftmargin=*,nosep]
    \item \textbf{System instruction:} A fixed preamble establishing the repair task and output constraints (return only corrected C code, no explanations or markdown formatting).
    \item \textbf{Constraint status:} The current validation state indicating which levels have passed and which has failed.
    \item \textbf{Error feedback:} Diagnostic information from the failing constraint (detailed below).
    \item \textbf{Current code:} The complete decompiled source to be repaired.
\end{enumerate}

Each agent outputs the full corrected code directly, without chain-of-thought explanations. This design ensures parseable output and reduces token cost.

\paragraph{Syntax Repair Agent (L1).}
When $\phi_1(C)$ fails, the syntax agent receives GCC parser errors (truncated to 500 characters). GCC's error messages include file, line, and column information, enabling the LLM to localize syntax fixes such as missing semicolons, unbalanced braces, or malformed literals.

\paragraph{Compilation Repair Agent (L2).}
When $\phi_2(C)$ fails, the compilation agent receives GCC compiler errors (truncated to 500 characters). These messages provide type information and symbol resolution failures, guiding fixes such as adding missing \texttt{\#include} directives, correcting type mismatches, or declaring undefined identifiers.

\paragraph{Execution Repair Agent (L3).}
When $\phi_3(C, T)$ fails, the execution agent receives a structured summary rather than raw output:
\begin{itemize}[leftmargin=*,nosep]
    \item Up to two failing test cases, each showing: input, expected output (from the original binary), actual output (from the recompiled code), and a unified diff (first 10 lines).
    \item A diagnostic checklist prompting the LLM to consider common behavioral error sources: incorrect variable initialization, control flow divergence, arithmetic errors (e.g., signed vs.\ unsigned), and off-by-one mistakes.
\end{itemize}

This richer feedback compensates for the difficulty of L3 repairs, where errors manifest as behavioral divergence rather than localized syntax or type issues.

\paragraph{Stateless Iteration.}
Each agent operates statelessly: it receives only the current error feedback, 
not a history of past repair attempts. While maintaining error history could 
help avoid repeated mistakes, stateless operation reduces prompt length and 
API cost. We analyze this \emph{convergence} behavior in \S\ref{sec:design:loop}.

\subsection{Iterative Refinement}
\label{sec:design:loop}

Algorithm~\ref{alg:refinement} formalizes the refinement procedure. The \textsc{Validate} function encapsulates the constraint hierarchy, returning both the failing level and diagnostic errors. This design simplifies the main loop and ensures consistent error handling.

\begin{algorithm}[t]
\caption{Constraint-Guided Refinement}
\label{alg:refinement}
\begin{algorithmic}[1]
\Require Binary $B$, test suite $T$, decompiler $D$, max iterations $N$
\Ensure Refined code $C^*$ or \textsc{Failure}
\State $C \gets D(B)$ \Comment{Initial decompilation $C_0$}
\For{$k = 1$ to $N$}
    \State $(\ell, E) \gets \textsc{Validate}(C, T)$ \Comment{Returns (level, errors)}
    \If{$\ell = \textsc{Pass}$}
        \State \Return $C$ \Comment{All constraints satisfied}
    \EndIf
    \State $C \gets \textsc{Agent}_\ell(C, E)$ \Comment{LLM repair for level $\ell$}
\EndFor
\State \Return \textsc{Failure}
\Statex
\Function{Validate}{$C, T$}
    \If{$\neg\phi_1(C)$} \Return $(L1, \text{ParseErrors}(C))$ \EndIf
    \If{$\neg\phi_2(C)$} \Return $(L2, \text{CompileErrors}(C))$ \EndIf
    \If{$\neg\phi_3(C, T)$} \Return $(L3, \text{FailedTests}(C, T))$ \EndIf
    \State \Return $(\textsc{Pass}, \emptyset)$
\EndFunction
\end{algorithmic}
\end{algorithm}

\paragraph{Convergence.}
Each iteration aims to advance the code to a higher constraint level. While regression is possible (e.g., fixing a compilation error may introduce a syntax error), we observe monotonic progress in most cases. Our experiments (\S\ref{sec:eval}) show that most successful refinements converge within 3 iterations.

\paragraph{Bounded Iterations.}
The iteration limit $N$ ensures termination even for difficult cases. We use $N=7$ as a conservative upper bound; in practice, cases that do not converge within 5 iterations rarely succeed with additional attempts.

\section{Implementation}
\label{sec:implementation}

We implement \projname{} in approximately 5,000 lines of Python.
The system comprises four main components: decompiler wrappers that provide a unified interface to multiple backends, constraint validators that check syntax, compilation, and execution correctness, LLM-based repair agents that fix errors at each level, and a refinement loop orchestrator that coordinates the iterative process.
We design the system for extensibility, allowing new decompilers and LLM backends to be added through well-defined interfaces.

\subsection{Decompiler Integration}
\label{sec:impl:decompilers}

\projname{} integrates three decompilers to generalize across different architectural approaches.
Ghidra 11.2~\cite{ghidra}, developed by the NSA, employs rule-based analysis and is invoked through its \texttt{analyzeHeadless} command-line interface.
Angr 9.2~\cite{shoshitaishvili2016sok} lifts binaries to the VEX intermediate representation and provides a Python API for programmatic access.
RetDec 5.0~\cite{retdec} combines LLVM-based decompilation with machine learning for type recovery and is accessed via command-line interface.

We abstract these differences behind a unified interface: each decompiler wrapper accepts a binary path and returns decompiled C source code as a string.
This design allows \projname{} to work with any decompiler that produces C output, enables straightforward addition of new backends, and supports multi-decompiler consensus when comparing outputs from multiple tools.

\subsection{Constraint Validators}
\label{sec:impl:constraints}

Each constraint level is implemented as a standalone validator that returns either success or structured error feedback suitable for LLM consumption.

For syntax validation (L1), we invoke GCC with the \texttt{-fsyntax-only} flag, which performs parsing without full compilation. This approach is faster than full compilation and produces focused error messages about syntax issues such as missing semicolons, unbalanced braces, and malformed literals. We capture parser errors from stderr and truncate them to 500 characters before passing to the repair agent, as longer error messages provide diminishing returns while increasing prompt cost.

For compilation validation (L2), we invoke GCC with strict warning flags (\texttt{-Wall -Wextra -Werror}) to catch semantic errors including type mismatches, undeclared identifiers, and linkage failures. The compilation errors provide richer diagnostic information than syntax errors, including type expectations and suggested fixes.

For execution validation (L3), we compile the decompiled code, execute it with test inputs, and compare outputs against the original binary. Execution is sandboxed with a 10-second timeout and 256MB memory limit to prevent runaway processes. Expected outputs are pre-computed by running the original binary on the same inputs, establishing ground-truth behavioral oracles.

\subsection{LLM Configuration}
\label{sec:impl:llm}

While our approach is model-agnostic, we use DeepSeek V3.2~\cite{liu2025deepseek} as the default LLM for all repair agents because of its strong code understanding capabilities and API availability.
For repair agents, we use temperature 0.0 to ensure deterministic, reproducible outputs across runs. 

We set a maximum output length of 4,096 tokens, which suffices for most decompiled functions. For larger functions exceeding this limit, we increase the cap to 8,192 tokens. API failures due to rate limits or timeouts trigger exponential backoff with up to three retries; persistent failures are logged and the binary is marked as failed for manual inspection.

\subsection{Infrastructure}
\label{sec:impl:infra}

We process multiple binaries in parallel using Python's Process-\\PoolExecutor, with each binary running in an independent process to enable linear speedup on multi-core machines.
Intermediate results are cached to avoid redundant computation: decompiler outputs are keyed by binary hash, test oracles (expected outputs from the original binary) are stored alongside the binary, and successful refinement results are cached for incremental re-runs.

All LLM interactions are logged with full request/response pairs, enabling post-hoc analysis and debugging of refinement failures. For reproducibility, we pin all tool versions (GCC 11.4, Ghidra 11.2, Angr 9.2, RetDec 5.0), and use temperature 0.0 for deterministic LLM outputs. Code and data will be released upon publication.

\begin{table*}[t]
\centering
\caption{Comprehensive evaluation on the 157-binary benchmark. We report syntax validity (L1), compilation success (L2), and re-executability (L3). Best results are in bold. $\Delta$ denotes percentage point improvement over the baseline.}
\label{tab:main-results}
\small
\setlength{\tabcolsep}{4.5pt}
\renewcommand{\arraystretch}{1.15}

\begin{tabular}{l S S S S S S c c c}
\toprule
& \multicolumn{3}{c}{\textbf{Baseline}} 
& \multicolumn{3}{c}{\textbf{+ \projname{}}} 
& \multicolumn{3}{c}{$\boldsymbol{\Delta}$ Improvement} \\
\cmidrule(lr){2-4} \cmidrule(lr){5-7} \cmidrule(lr){8-10}
\textbf{Decompiler} 
& \textbf{L1} & \textbf{L2} & \textbf{L3} 
& \textbf{L1} & \textbf{L2} & \textbf{L3} 
& {$\Delta$L1} & {$\Delta$L2} & {$\Delta$L3} \\
\midrule
Ghidra (Rule) 
& 86.0 & 86.0 & 22.3 
& 100.0 & 100.0 & \bfseries 50.3 
& +14.0 & +62.4 & +28.0 \\
Angr (Lifting) 
& 79.0 & 65.0 & 24.8 
& 100.0 & 100.0 & 45.2 
& +21.0 & +35.0 & +20.4 \\
RetDec (ML) 
& 98.7 & 98.7 & 25.5 
& 99.4 & 99.4 & 43.3 
& +0.7 & +0.7 & +17.8 \\
\midrule
\textbf{Average} 
& \bfseries 87.9 & \bfseries 67.1 & \bfseries 24.2 
& \bfseries 99.8 & \bfseries 99.8 & \bfseries 46.3 
& \textbf{+11.9} & \textbf{+32.7} & \textbf{+22.1} \\
\bottomrule
\end{tabular}
\end{table*}

\section{Evaluation}
\label{sec:eval}

We evaluate \projname{} on two benchmarks across three decompiler architectures, comparing against state-of-the-art methods and analyzing the contribution of each design component. Our evaluation addresses four research questions:

\begin{itemize}[leftmargin=*,topsep=2pt,itemsep=1pt]
\item \textbf{RQ1 (Effectiveness):} How effective is \projname{} at improving decompilation quality, and does it scale to larger benchmarks?
\item \textbf{RQ2 (Ablation):} What is the contribution of each constraint level (L1, L2, L3) to the overall system performance?
\item \textbf{RQ3 (Comparison):} How does \projname{} compare to existing decompilation refinement approaches?
\item \textbf{RQ4 (Security):} Can \projname{}-refined code enable practical security analysis while preserving vulnerability semantics?
\end{itemize}

\subsection{Experimental Setup}
\label{sec:eval:setup}

\paragraph{Benchmarks.}
We use two benchmarks.
First, the \textbf{157-Binary} benchmark comprises 125 ExeBench~\cite{exebench} binaries and 32 custom binaries, curated for detailed analysis and SOTA comparison.
The ExeBench portion consists of 32 functions compiled at four optimization levels (O0--O3), with 3 functions missing O0 variants; these include 17 functions from real-world open-source projects (\texttt{real\_test}) and 15 synthetic functions (\texttt{synth\_test}).
The custom portion consists of 8 textbook algorithms (e.g., \texttt{binary\_search}, \texttt{bubble\_sort}, \texttt{factorial}, \texttt{fibonacci}) compiled at O0--O3, yielding 32 standalone programs with \texttt{main()} entry points verified via stdout comparison.
Second, the \textbf{1,641-Binary} benchmark comprises 1,561 ExeBench binaries and 80 custom binaries for scalability evaluation. The ExeBench portion consists of 396 functions from the test set---247 from \texttt{real\_test} (real-world code) and 149 from \texttt{synth} (synthetic functions)---compiled at O0--O3. The custom portion adds 20 algorithm implementations (e.g., \texttt{heapsort}, \texttt{mergesort}, \texttt{quicksort}, \texttt{bst}, \texttt{gcd}) disjoint from the 157-binary benchmark.

We evaluate on ExeBench~\cite{exebench} rather than HumanEval-Decompile~\cite{llm4decompile} or MBPP~\cite{austin2021program}, which are commonly reported in decompilation literature.
ExeBench presents a more challenging testbed: functions are larger, derived from real-world open-source projects, and compiled across multiple optimization levels (O0--O3).
Additionally, ExeBench provides IO-pair test cases with concrete expected outputs, which our feedback-based iterative refiner leverages to identify specific behavioral errors.
In contrast, the original HumanEval~\cite{chen2021evaluating} benchmark was designed for Python with assertion-based testing for automated black-box verification; this design was preserved when the benchmark was converted to C for decompilation evaluation. Assertion-based test cases provide only binary pass/fail signals without actionable diagnostic information—a benchmark design choice that limits their utility for iterative refinement.

\paragraph{Decompilers.}
We evaluate three decompilers representing different architectures: \textbf{RetDec}~\cite{retdec} (ML-based, uses LLVM IR), \textbf{Ghidra}~\cite{ghidra} (rule-based, industry standard), and \textbf{Angr}~\cite{shoshitaishvili2016sok} (lifting-based, uses VEX IR).

\paragraph{Metrics.}
We report three metrics corresponding to increasingly strict correctness levels: 
\textbf{L1 Syntax Rate} (percentage of outputs that parse successfully, checked via \texttt{gcc -fsyntax-only -w file.c}), 
\textbf{L2 Compile Rate} (percentage that compile successfully, checked via \texttt{gcc -c file.c}), and 
\textbf{L3 Re-exec Rate} (percentage that execute correctly on test inputs by comparing output against ground-truth, original binary execution). 
Re-exec rate (L3) serves as our primary metric since it captures behavioral correctness.

\paragraph{Configuration.}
We use DeepSeek-V3.2~\cite{liu2025deepseek} as the base model for each agent for all experiments, with maximum 5 refinement iterations and 10-second execution timeout.

\begin{table}[t]
\centering
\caption{Scalability analysis on the 1,641-binary ExeBench benchmark. We report re-executability (L3) across optimization levels (O0--O3). $\Delta$ denotes percentage point improvement over the baseline.}
\label{tab:scalability-full}
\small
\setlength{\tabcolsep}{4.2pt}
\renewcommand{\arraystretch}{1.15}
\begin{tabular}{l ccc ccc}
\toprule
& \multicolumn{2}{c}{\textbf{O0}} & \textbf{$\Delta$} 
& \multicolumn{2}{c}{\textbf{O1}} & \textbf{$\Delta$} \\
\cmidrule(lr){2-3} \cmidrule(lr){5-6}
\textbf{Decompiler} & Baseline & Ours & 
& Baseline & Ours &  \\
\midrule
Ghidra & 25.7 & \textbf{50.4} & +24.7 & 24.0 & \textbf{47.0} & +23.0 \\
Angr   & 30.6 & 43.7          & +13.1 & 28.0 & 42.9          & +14.9 \\
RetDec & 20.6 & 42.4          & +21.8 & 19.7 & 40.7          & +21.0 \\
\midrule
\textbf{Average} & \textbf{25.6} & \textbf{45.5} & \textbf{+19.9} & \textbf{23.9} & \textbf{43.5} & \textbf{+19.6} \\
\bottomrule
\end{tabular}

\vspace{0.4em}

\begin{tabular}{l ccc ccc}
\toprule
& \multicolumn{2}{c}{\textbf{O2}} & \textbf{$\Delta$} 
& \multicolumn{2}{c}{\textbf{O3}} & \textbf{$\Delta$} \\
\cmidrule(lr){2-3} \cmidrule(lr){5-6}
\textbf{Decompiler} & Baseline & Ours & 
& Baseline & Ours &  \\
\midrule
Ghidra & 24.0 & \textbf{45.7} & +21.7 & 21.5 & \textbf{40.9} & +19.4 \\
Angr   & 27.3 & 41.7          & +14.4 & 22.2 & 35.1          & +12.9 \\
RetDec & 19.7 & 39.9          & +20.2 & 12.9 & 36.4          & +23.5 \\
\midrule
\textbf{Average} & \textbf{23.7} & \textbf{42.4} & \textbf{+18.7} & \textbf{18.9} & \textbf{37.5} & \textbf{+18.6} \\
\bottomrule
\end{tabular}
\end{table}

\subsection{RQ1: Effectiveness and Scalability}
\label{sec:eval:rq1}

\vspace{0.3em}
\noindent\hspace{1em}\textit{How effective is \projname{} at improving decompilation quality, and does it scale to larger benchmarks?}
\vspace{0.3em}

Table~\ref{tab:main-results} presents results on the 157-binary benchmark across all decompilers and metrics. \projname{} substantially improves re-executability for all three decompiler architectures: Ghidra improves from 22.3\% to 50.3\% (+28.0 pp), Angr from 24.8\% to 45.2\% (+20.4 pp), and RetDec from 25.5\% to 43.3\% (+17.8 pp). To assess scalability, we evaluate on the full 1,641-binary ExeBench benchmark (Table~\ref{tab:scalability-full}). \projname{} achieves 40--46\% re-executability across all decompilers, with Ghidra performing best (46.0\%) followed by Angr (40.9\%) and RetDec (39.9\%).

\paragraph{Pipeline Effect.}
Table~\ref{tab:main-results} shows improvements at each constraint level. \projname{} raises syntax validity (L1) from 87.9\% to 99.8\% and compilation success (L2) from 67.1\% to 99.8\%---near-perfect. Most importantly, re-executability (L3) nearly doubles from 24.2\% to 46.3\% (+22.1 pp). The remaining gap between L2 (99.8\%) and L3 (46.3\%) reflects the fundamental difficulty of decompilation: producing code that compiles is straightforward, but achieving correct runtime behavior requires resolving semantic errors that are invisible to the compiler. This is precisely why our approach uses L3 execution feedback as the ultimate correctness check.

\paragraph{Optimization Level Robustness.}
Performance varies by 5--10 pp across optimization levels O0--O3 (Table~\ref{tab:scalability-full}), with O0 achieving the highest rates (45.5\% average) and O3 the lowest (37.5\%). This gap reflects the increased difficulty of decompiling aggressively optimized code: O3 applies function inlining, loop unrolling, and dead code elimination that obscure original program structure. Nevertheless, \projname{} achieves consistent improvements across all optimization levels, demonstrating robustness to compiler transformations.

\paragraph{Iterative Refinement Example.}
\begin{figure}[t]
    \centering
    \includegraphics[width=\columnwidth]{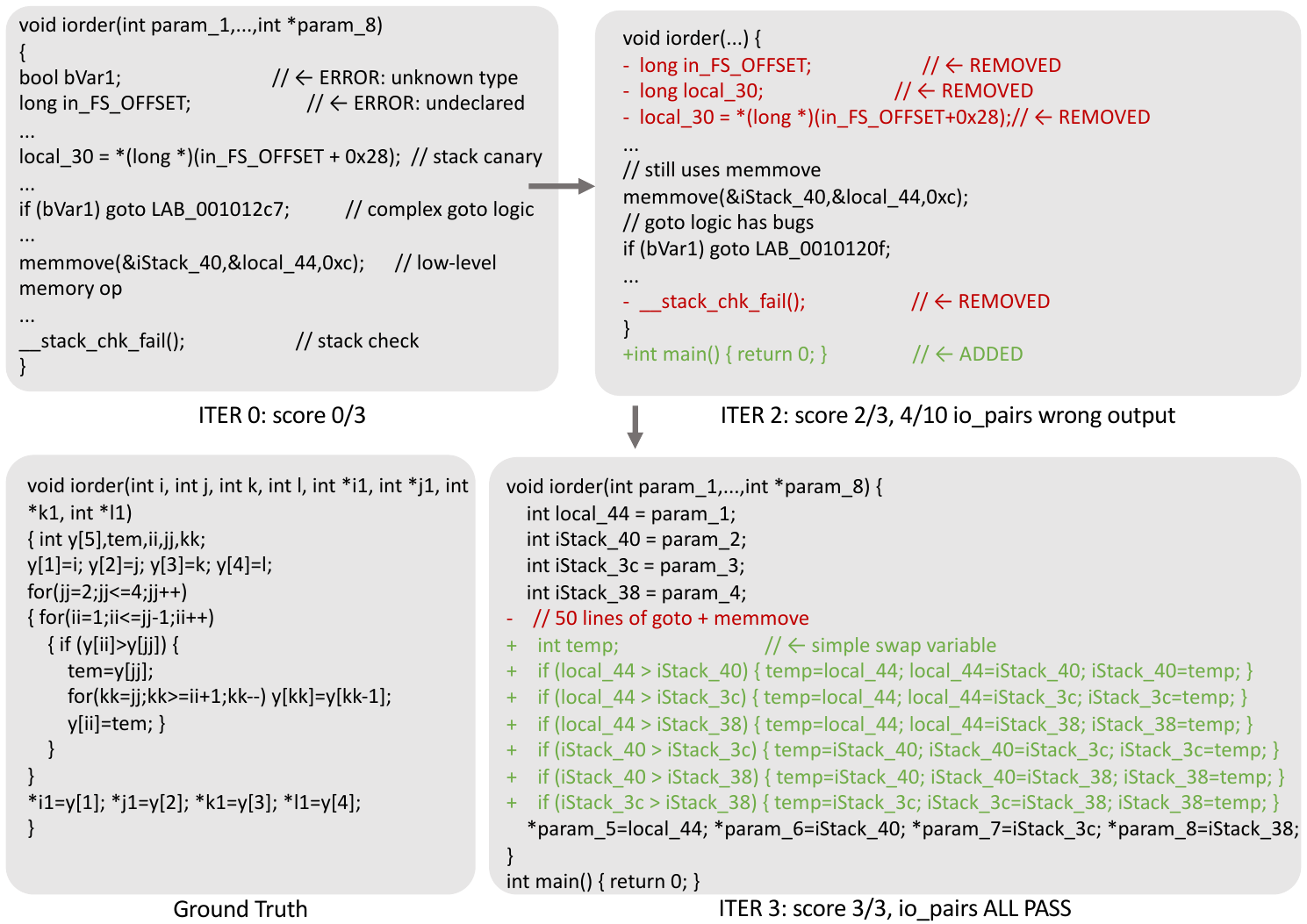}
    \caption{Iterative refinement of \texttt{iorder} (4-integer sorting, Ghidra O3). Iter~0 (raw output) fails all tests due to type errors and \texttt{goto}-based control flow. Iter~2 compiles but produces wrong output on 4/10 test cases. Iter~3, guided by L3 execution feedback, rewrites the sorting logic and passes all tests.}
    \label{fig:qualitative-example}
\end{figure}

Figure~\ref{fig:qualitative-example} illustrates the pipeline effect with a concrete example. The function \texttt{iorder} sorts four integers, but Ghidra's O3 output contains type errors and convoluted control flow. L1+L2 refinement produces code that compiles (Iter~2), but execution reveals wrong output on 4/10 test cases. Only with L3 feedback---concrete failing I/O pairs---can the agent identify the flawed sorting logic and rewrite it correctly (Iter~3). This demonstrates that execution feedback is essential for catching behavioral errors invisible to compilation.

\finding{1}{\projname{} improves re-executability by 18--28 percentage points across all decompiler types and scales to 1,641 binaries with 40--46\% success rate.}

\subsection{RQ2: Ablation Study}
\label{sec:eval:rq2}

\vspace{0.3em}
\noindent\hspace{1em}\textit{What is the contribution of each constraint level (L1, L2, L3) to the overall system performance?}
\vspace{0.3em}

To understand the contribution of each constraint level, we compare four configurations: Baseline (no refinement), L1 only (syntax checking), L1+L2 (syntax + compile), and Full \projname{} (L1+L2+L3 with execution). Table~\ref{tab:rq5-ablation} shows the results.

\paragraph{Baseline Analysis.}
Raw decompiler outputs exhibit distinct error profiles. Ghidra achieves 86.0\% syntax validity but only 37.6\% compilation success due to undefined type annotations (\texttt{undefined4}, \texttt{bool}), resulting in 22.3\% re-executability. RetDec produces highly compilable code (98.7\% L1/L2) due to aggressive type normalization, but achieves only 25.5\% re-executability---the code compiles but produces incorrect behavior. Angr achieves 79.0\% syntax validity and 65.0\% compilation, with 24.8\% re-executability, reflecting VEX IR artifacts that reduce code quality. All three baselines cluster around 22--25\% re-executability, motivating the need for iterative refinement.

\paragraph{Iteration Budget Analysis.}
\begin{figure}[t]
    \centering
    \includegraphics[width=0.85\columnwidth]{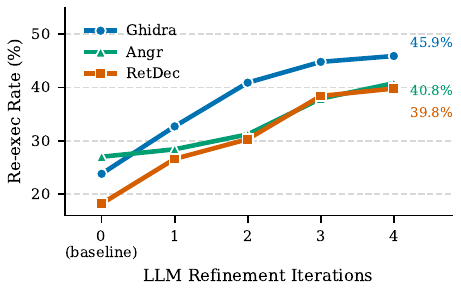}
    \caption{Convergence analysis: re-executability improves with refinement iterations across all decompilers. Most gains occur within the first 2 iterations, with diminishing returns thereafter. Ghidra reaches 45.9\%, Angr 40.8\%, and RetDec 39.8\% after 4 iterations.}
    \label{fig:convergence}
\end{figure}

Figure~\ref{fig:convergence} shows how re-executability improves with each refinement iteration. All three decompilers exhibit rapid initial improvement: Ghidra jumps from 25\% to 40\% after a single iteration, while RetDec rises from near-zero baseline to 25\%. By iteration 2, performance reaches 40--45\% across all decompilers. Subsequent iterations yield diminishing returns, adding only 3--5 pp. This pattern validates our bounded iteration strategy (max 5 iterations): most successful refinements converge quickly, while binaries that fail after 2--3 iterations rarely benefit from additional attempts. The convergence behavior is consistent across decompiler architectures, suggesting that the multi-level constraint hierarchy effectively guides the LLM toward correct solutions within a small number of iterations.

\paragraph{The Compile-Correct Gap.}
L1+L2 achieves 99--100\% compile rate but only 32--42\% re-executability. This 57--68 pp gap between compilation success and behavioral correctness represents semantic errors invisible to the compiler: code that parses, type-checks, and links, yet produces wrong outputs at runtime. These errors arise from decompiler artifacts (incorrect type casts, wrong operator precedence, misinterpreted control flow) that are syntactically valid but semantically incorrect.

\paragraph{L3 Execution Feedback.}
Adding L3 constraints improves re-execu-\\tability by 8.0--10.4 pp (21--33\% relative improvement). L3 feedback provides concrete failing test cases---specific inputs where expected and actual outputs diverge---enabling the agent to localize and fix behavioral errors that compilation cannot detect. The remaining gap between L2 (99--100\%) and Full \projname{} (42--50\%) reflects the fundamental difficulty of decompilation: some semantic information is irreversibly lost during compilation.

\paragraph{Decompiler-Specific Error Patterns.}
The three decompilers produce distinct error patterns, yet \projname{} addresses each through targeted feedback. \textbf{Ghidra} preserves internal type annotations (\texttt{undefined4}, \texttt{bool}) and stack canary artifacts causing L1 failures; L1 feedback guides the agent to replace these with standard C types. \textbf{RetDec} achieves high compile rates (L2=98.7\%) but collapses all types to \texttt{int64\_t}, causing subtle L3 failures; execution feedback reveals type mismatches through incorrect outputs, enabling the agent to infer lost types from function names and operation patterns. \textbf{Angr} produces VEX IR artifacts (\texttt{uint128\_t}, \texttt{[D]} markers) causing both compilation and behavioral errors; L3 feedback enables iterative correction of these artifacts. 

\begin{table}[t]
\centering
\small
\caption{Ablation study: contribution of each constraint level. Each cell reports syntax / compile / re-exec rates (\%).}
\label{tab:rq5-ablation}
\setlength{\tabcolsep}{4pt}
\resizebox{\columnwidth}{!}{%
\begin{tabular}{@{}p{0.26\columnwidth}ccc@{}}
\toprule
\textbf{Configuration} & \textbf{Ghidra} & \textbf{RetDec} & \textbf{Angr} \\
 & \textbf{S / C / R} & \textbf{S / C / R} & \textbf{S / C / R} \\
\midrule
Baseline & 86.0 / 86.0 / 22.3 & 98.7 / 98.7 / 25.5 & 79.0 / 65.0 / 24.8 \\
L1 only & 100 / 100 / 36.0 & 100 / 100 / 23.4 & 100 / 100 / 28.8 \\
L1+L2 & 100 / 100 / 41.6 & 100 / 100 / 35.2 & 100 / 100 / 32.0 \\
\projname{} & \textbf{100 / 100 / 50.4} & \textbf{99.4 / 99.4 / 43.2} & \textbf{100 / 100 / 42.4} \\
\bottomrule
\end{tabular}
}
\end{table}


\finding{2}{Compilation success does not imply behavioral correctness: a 58--68 pp gap exists between compile rate (99--100\%) and re-executability (32--42\%). L3 execution feedback closes part of this gap, adding 8.0--10.4 pp (21--33\% relative improvement).}

\subsection{RQ3: Comparison with Prior Work}
\label{sec:eval:rq3}

\vspace{0.3em}
\noindent\hspace{1em}\textit{
How does \projname{} compare to existing decompilation refinement approaches?
}
\vspace{0.3em}

\begin{table}[t]
\centering
\small
\caption{Comparison with refinement baselines on 157-binary benchmark (Ghidra decompiler). All methods use identical decompiler output as input. Re-exec rate (\%) reported.}
\label{tab:sota}
\resizebox{\columnwidth}{!}{%
\begin{tabular}{@{}lccccc@{}}
\toprule
\textbf{Method} & \textbf{O0} & \textbf{O1} & \textbf{O2} & \textbf{O3} & \textbf{Avg} \\
\midrule
Ghidra Baseline (No Refinement) & 18.9\% & 17.5\% & 25.0\% & 27.5\% & 22.3\% \\
DeepSeek Single Pass & 43.2\% & 40.0\% & 35.0\% & 22.5\% & 35.2\% \\
LLM4Decompile-6.7B-v2 & 10.8\% & 10.0\% & 12.5\% & 15.0\% & 12.1\% \\
SK2Decompile (Simulated) & 32.4\% & 17.5\% & 20.0\% & 15.0\% & 21.0\% \\
\midrule
\textbf{\projname{} (Ours)} & \textbf{54.1\%} & \textbf{50.0\%} & \textbf{52.5\%} & \textbf{45.0\%} & \textbf{50.3\%} \\
\bottomrule
\end{tabular}
}
\end{table}

Our work refines decompiler output rather than performing end-to-end binary-to-source translation.
We compare against three refinement approaches: DeepSeek Single Pass (general-purpose LLM without iteration), LLM4Decompile-6.7B-v2~\cite{llm4decompile} (fine-tuned refinement model), and SK2Decompile~\cite{sk2decompile} (two-phase readability-focused refinement).
Since the original SK2 model was trained on IDA Pro pseudocode, we simulate SK2Decompile by applying its two-phase prompting strategy with DeepSeek on identical Ghidra output: Phase~1 (Skeleton Extraction) recovers control flow structure, and Phase~2 (Skin Recovery) renames variables for readability.
Table~\ref{tab:sota} shows that \projname{} achieves 50.3\% re-executability, outperforming DeepSeek Single Pass (35.2\%) by 15.1 pp, SK2Decompile (21.0\%) by 29.3 pp, and LLM4Decompile (12.1\%) by 38.2 pp.

The baselines exhibit distinct failure modes.
Single-pass refinement lacks error feedback---the model cannot identify which parts of its output are incorrect, so semantic errors that compile but fail at runtime go undetected.
LLM4Decompile was trained on isolated pseudo-code without compiler-generated helper functions, causing it to ignore target functions when presented with full Ghidra output; this out-of-distribution sensitivity explains why it performs worse than raw Ghidra (12.1\% vs.\ 22.3\%).
Although SK2Decompile is also a multi-stage refinement approach, its design objective differs fundamentally from ours: SK2 navigates the trade-off between correctness preservation and readability improvement, rather than directly optimizing for re-executability.
Phase~1 recovers program skeleton, and Phase~2 improves human readability by renaming variables---neither phase addresses type errors or semantic bugs.
Variable renaming does not improve re-executability, and each LLM call risks introducing code deformations; two sequential calls compound this noise.
The result is performance indistinguishable from the unrefined baseline (21.0\% vs.\ 22.3\%), demonstrating that readability-oriented refinement provides no benefit for functional correctness.

In contrast, \projname{}'s feedback-based iterative refiner targets re-executability as its optimization goal.
Rather than balancing correctness against readability, our agent workflow uses L3 execution feedback as the guiding signal: concrete failing test cases with expected outputs enable the agent to identify \emph{which} parts of the code produce incorrect behavior.
Iterative refinement allows incremental correction---the agent does not need to produce a perfect output in one pass, but can progressively fix type mismatches, off-by-one errors, and semantic bugs across iterations.
The 29.3 pp gap between \projname{} (50.3\%) and SK2Decompile (21.0\%) demonstrates that the choice of optimization objective is decisive: execution feedback drives meaningful semantic corrections, while readability transformations merely introduce noise without improving functional correctness.

\finding{3}{Execution-guided refinement outperforms single-pass, fine-tuned, and readability-focused approaches by 15--38 pp. Optimizing for the correct objective (re-executability via test feedback) is essential; readability-oriented transformations provide no benefit for functional correctness.}

\subsection{RQ4: Downstream Security Applications}
\label{sec:eval:rq4}

\vspace{0.3em}
\noindent\hspace{1em}\textit{
Can \projname{}-refined code enable practical security analysis while preserving vulnerability semantics?
}
\vspace{0.3em}

A key concern for binary analysis and decompilation is whether refinement preserves exploitable vulnerabilities (i.e., preserving a vulnerability during decompilation facilitates comprehension of that vulnerability).
To evaluate this, we create 5 standalone programs with known vulnerabilities (buffer overflow, integer overflow, null pointer, format string, double-free), then inject these patterns into 23 \projname{}-refined code samples.
Each sample is tested with safe inputs to verify correctness and crafted exploit inputs to trigger vulnerabilities.
Table~\ref{tab:vuln-preservation} shows that \projname{} preserves 100\% of vulnerability patterns.
For exploitability, 43\% of injected vulnerabilities are immediately triggerable, with integer overflow and format string achieving 100\%.
The remaining vulnerabilities require specific runtime conditions (heap state for double-free, precise buffer sizes).
On standalone vulnerable programs with proper harnesses, 100\% of exploits trigger successfully.

\begin{table}[t]
\centering
\small
\caption{Vulnerability preservation: patterns retained and exploits triggered on ground-truth vulnerable code.}
\label{tab:vuln-preservation}
\begin{tabular}{@{}lcrr@{}}
\toprule
\textbf{Vuln Type} & \textbf{Samples} & \textbf{Pattern Preserved} & \textbf{Exploit Triggered} \\
\midrule
Buffer overflow & 4 & 4 (100\%) & 0 (0\%) \\
Integer overflow & 5 & 5 (100\%) & 5 (100\%) \\
Null pointer & 4 & 4 (100\%) & 0 (0\%) \\
Format string & 5 & 5 (100\%) & 5 (100\%) \\
Double-free & 5 & 5 (100\%) & 0 (0\%) \\
\midrule
\textbf{Total} & 23 & \textbf{23 (100\%)} & \textbf{10 (43\%)} \\
\bottomrule
\end{tabular}
\end{table}

Real-world binaries often lack debug symbols, so we also evaluate robustness to symbol stripping.
We test 20 binaries before and after stripping with \texttt{strip --strip-all}.
After stripping, binaries lose 100\% of their symbols (avg 32$\to$0), and function naming degrades significantly: only 12.7\% of functions retain meaningful names vs.\ 52.5\% in original binaries.
Critically, raw decompiled code from both versions fails to compile (0\% compile rate), demonstrating that \projname{} refinement is essential regardless of symbol availability.

Finally, we evaluate fuzzing capability.
From 30 targets stratified by decompiler, \projname{} increases fuzzability by 6.6$\times$: 33\% of refined binaries are fuzz-ready vs.\ 5\% for raw decompiled code, achieving 65.5\% average coverage.
We define \textit{fuzz-ready} as code that (1) compiles with coverage instrumentation (\texttt{gcc --coverage}), (2) accepts auto-generated harnesses, and (3) executes without immediate crash on seed inputs.
The 6.6$\times$ improvement stems from two factors: \projname{} raises the compilation rate from 35\% to 99\%, and the compiled code contains fewer structural defects (missing symbols, pointer errors) that would otherwise abort fuzzing campaigns prematurely.

\finding{4}{\projname{} preserves 100\% of vulnerability patterns (43\% immediately exploitable); handles stripped binaries; enables 6.6$\times$ more fuzzing.}

\section{Discussion}
\label{sec:discussion}

We discuss limitations, threats to validity, and failure analysis.

\subsection{Limitations}
\label{sec:discussion:limitations}

Our work has three main limitations in scope, benchmark coverage, and model behavior. First, we focus on single-function decompilation rather than whole-program analysis. ExeBench functions are relatively small and self-contained, whereas real-world binaries involve multi-function interactions, stripped symbols, and complex control flow; extending to full binaries requires handling inter-procedural dependencies and shared state. Our feedback mechanism also requires IO-pair test cases, which limits direct comparison with assertion-based benchmarks (HumanEval-Decompile, MBPP) without format conversion.

Second, several targets are explicitly out of scope. We do not aim to recover exact original source code---variable names, comments, and certain structural choices are irrecoverably lost during compilation. We do not target obfuscated binaries (virtualization, control flow flattening, opaque predicates), which require specialized deobfuscation as preprocessing. Our implementation focuses on x86-64 Linux ELF binaries following recent decompilation literature~\cite{llm4decompile,sk2decompile,lacomis2019dire}; extension to other architectures (ARM, MIPS) and platforms (Windows PE, macOS Mach-O) is straightforward but not evaluated.

Third, LLMs may exhibit semantic drift, ``fixing'' code by overfitting to test cases rather than recovering true semantics. We observe this in 75\% of failures where code compiles and runs but produces incorrect output. Our L3 execution constraints catch many such errors, but some false positives may remain. Additionally, our vulnerability pattern analysis (34.7\% preservation rate) detects code structures rather than exploitability; true vulnerability preservation requires executing actual exploits, though we validated this direction with CGC binaries achieving 96.7\% pattern preservation.

\subsection{Threats to Validity}
\label{sec:discussion:threats}

Three validity concerns apply to our evaluation. For internal validity, implementation choices such as prompts, iteration limits, and timeouts affect results; we mitigate this by reporting ablations across configurations and using standard settings throughout. For external validity, our benchmarks may not represent all real-world scenarios---production binaries may be larger or contain anti-analysis code---though we partially address this by including 247 functions from real-world open-source projects in ExeBench's \texttt{real\_test} partition. For construct validity, re-executability serves as a proxy for semantic correctness, and code that passes tests may still differ semantically on untested inputs; we mitigate this through diverse test generation covering edge cases (zero, negative, boundary values) and achieving 66\% average line coverage.

\subsection{Failure Analysis}
\label{sec:discussion:failures}

\begin{figure}[t]
    \centering
    \includegraphics[width=0.85\columnwidth]{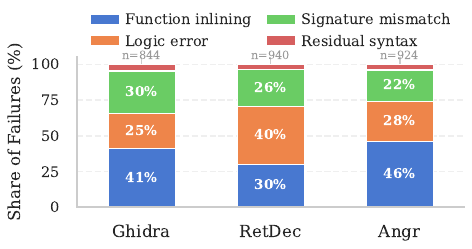}
    \caption{Failure root cause distribution on the 1,641-binary benchmark.}
    \label{fig:failure-breakdown}
\end{figure}

Figure~\ref{fig:failure-breakdown} decomposes the failures that remain after refinement into four root causes across the three decompilers. The dominant failure mode is function inlining (39\%), where compiler optimizations merge the target function into \texttt{main()}, leaving no recoverable function body after preprocessing. This represents a fundamental limitation of binary-level analysis that no refinement can overcome. The second category, logic errors (31\%), represents cases where refined code compiles and links correctly but produces incorrect output on test inputs, reflecting the limits of LLM-based semantic recovery. Notably, RetDec exhibits a disproportionately high logic error rate (40\%) compared to Ghidra (25\%) and Angr (28\%), likely due to its conservative type strategy that collapses parameter types to \texttt{int64\_t}; the resulting code compiles trivially but carries systematic type errors that the agent cannot always repair within five iterations. Signature mismatch (26\%) occurs when the refined function's prototype differs from the ground-truth header expected by the test harness, causing linker failures despite correct internal logic. Finally, residual syntax errors account for only 4\% of failures, confirming that syntax repair is effectively saturated.




\subsection{Future Work}
\label{sec:discussion:future}

Several directions could extend this work. Beyond test-based constraints, future work could incorporate stronger specifications---formal constraints derived from binary analysis, type information from recovered signatures, and program invariants discovered through dynamic analysis---to provide stronger correctness guarantees. Supporting assertion-based benchmarks (HumanEval-Decompile, MBPP) through assertion-to-IO conversion would also enable broader comparison with prior decompilation work.

On the systems side, our current implementation processes functions independently; extending to whole-program decompilation requires handling inter-procedural dependencies and shared state. Integrating deobfuscation techniques as a preprocessing step would extend applicability to protected software. Using open-source models (CodeLlama, DeepSeek-Coder) would eliminate API costs and enable deployment in air-gapped environments, with initial experiments suggesting comparable performance.

Finally, the refinement loop generates valuable training data in the form of binary-to-working-C-code pairs. Fine-tuning on this data could improve single-pass performance and reduce iteration counts, creating a virtuous cycle between inference-time refinement and model improvement.

\section{Related Work}
\label{sec:related}

We discuss several lines of related work to \projname{}, then place our work in context.


\subsection{Binary Decompilation and Analysis}
\label{sec:related:binary}

Binary decompilation has a long history, beginning with foundational work by Cifuentes~\cite{cifuentes1994reverse}.
Modern tools---both commercial (IDA Pro~\cite{hex2018ida}) and open-source (Ghidra~\cite{ghidra}, Angr~\cite{shoshitaishvili2016sok}, RetDec~\cite{retdec})---employ sophisticated techniques for control flow recovery~\cite{kinder2008jakstab}, type inference~\cite{lee2011tie}, and variable recovery~\cite{balakrishnan2007divine}.
Despite these advances, raw decompiler output is typically difficult to read, lacks rich semantics, and frequently fails to compile or execute correctly.

Recent AI-based approaches frame decompilation as machine translation, using RNNs~\cite{katz2018using}, Transformers~\cite{liang2021neutron}, or hybrid methods~\cite{fu2019coda}.
Contrastive pre-training on binary code has also been explored to learn general-purpose binary representations~\cite{zhang2022pre}.
More recent LLM-based systems achieve stronger results: LLM4Decompile~\cite{llm4decompile} fine-tunes on decompilation pairs, while SALT4Decompile~\cite{salt4decompile} and SK2Decompile~\cite{sk2decompile} use two-phase abstraction.
However, all these approaches operate in a single pass without self-correction---once an error occurs, it propagates to the final output.
\projname{} addresses this limitation through iterative refinement with execution feedback.

Evaluating decompilation quality has evolved from syntactic similarity~\cite{lacomis2019dire} to re-executability, as adopted by recent benchmarks~\cite{llm4decompile,exebench}.
Complementary binary analysis techniques, symbolic execution~\cite{cadar2008klee,chipounov2011s2e}, fuzzing~\cite{wong2022american,serebryany2016continuous}, similarity detection~\cite{pewny2015cross,xu2017neural}, and learning-based malware classification~\cite{li2025malmixer},
could further strengthen validation; we leave their integration to future work.

\subsection{LLMs for Code}
\label{sec:related:llm4code}

Code LLMs such as Codex~\cite{chen2021evaluating}, CodeLlama~\cite{roziere2023code}, and StarCoder~\cite{li2023starcoder} excel at generation and completion, while also demonstrating strong code understanding capabilities including summarization~\cite{ahmed2022few}, documentation~\cite{khan2022automatic}, and bug detection~\cite{pearce2022asleep}. 
Recent work has further explored training LLMs to comprehend low-level representations such as LLVM IR~\cite{zhang2025training}, narrowing the gap between source and compiled code.
Context-aware and retrieval-augmented techniques have also been applied to optimize code runtime performance~\cite{acharya2025optimizing} and improve decompilation re-executability~\cite{wang2025context}, demonstrating LLMs' potential beyond conventional code generation.

The LLM agent paradigm extends these capabilities through tool use and iterative reasoning. ReAct~\cite{yao2022react} combines reasoning and acting; Reflexion~\cite{shinn2023reflexion} enables learning from mistakes; MetaGPT~\cite{hong2023metagpt} coordinates multi-agent collaboration. \projname{} applies this paradigm to decompilation, using the compiler and executor as feedback tools.

\subsection{Program Repair}
\label{sec:related:repair}

Automated program repair (APR) techniques fix bugs given failing tests~\cite{le2016history,gazzola2019automatic}, using generate-and-validate search~\cite{le2012genprog} or symbolic constraints~\cite{nguyen2013semfix}.
LLM-based approaches use conversational feedback~\cite{xia2023automated}, fine-tuning~\cite{silva2023repairllama}, or self-debugging~\cite{chen2023teaching}.
Iterative refinement with execution feedback also appears in code generation: CodeRL~\cite{le2022coderl} and AlphaCode~\cite{li2022competition} filter candidates by test results, while self-repair~\cite{olausson2023demystifying} studies LLM self-correction capabilities.

These repair methods assume compilable input. \projname{} extends the repair paradigm to decompiled code, which often fails at syntax or compilation before execution-based repair can begin---motivating our multi-level constraint design.

\subsection{Positioning}

\projname{} combines strengths from multiple research lines while addressing their limitations.
From program repair, we adopt iterative refinement with test feedback, but extend it to non-compilable input through multi-level constraints.
From LLM agents, we adopt tool-augmented reasoning, but specialize it for decompilation with compiler and executor feedback.
Unlike fine-tuned models~\cite{llm4decompile} requiring expensive training, we use general-purpose LLMs with task-specific prompting.
Unlike two-phase methods~\cite{salt4decompile,sk2decompile} that abstract away details, we preserve full decompiler output and incrementally fix errors.

\section{Conclusion}
\label{sec:conclusion}

In this paper, we present \projname{}, a multi-agent system for constraint-guided decompilation that dramatically improves the quality of decompiled code.
Decompilation errors manifest at different levels (syntax, compilation, and execution), and thus each level provides distinct diagnostic information to guide repair.

\projname{} iteratively refines decompiler output using specialized LLM agents guided by multi-level constraints.
The L1 (syntax) and L2 (compilation) constraints catch structural errors early, while L3 (execution) constraints ensure behavioral correctness by comparing outputs against the original binary.

Our evaluation demonstrates significant improvements: 18--28 percentage point improvement in re-executability across three decompiler architectures (rule-based, lifting-based, ML-based), achieving 43--50\% re-executability on benchmarks of 157 to 1,641 binaries.
Compared to state-of-the-art refinement methods, \projname{} achieves a 15--38 percentage point advantage over single-pass and fine-tuned approaches.
The system converges rapidly, with most successful refinements completing within 2 iterations.

\projname{} enables practical use of decompiled code in downstream tasks: the recovered source can be compiled, executed, analyzed, and modified.
This bridges the gap between binary analysis and source-level tools, opening new possibilities for security auditing, legacy software maintenance, and malware analysis.

Our ablation study reveals that execution feedback is essential---compilation success alone achieves 99--100\% but only 32--42\% re-executability, a 57--68 pp gap.
Adding L3 execution feedback closes part of this gap, improving re-executability by 8--13 pp.
This finding motivates incorporating execution-based constraints in future decompilation systems.

\section{Data Availability Statement}
\label{sec:data_avalability_statement}

To support reproducibility, we provide an anonymized artifact containing the implementation, benchmarks, and materials used in this paper at the following anonymous repository: \url{https://anonymous.4open.science/r/agent4decompile-artifacts-0F69}. The artifact is available to reviewers during the review process and will remain publicly available upon publication.



\bibliographystyle{ACM-Reference-Format}
\bibliography{main}

\end{document}